\documentclass[epj]{svjour}
\usepackage{graphics}
\usepackage{amsmath, amssymb}
\usepackage{epsf}
\usepackage{rotating}
\usepackage{graphicx,epsfig}% Include figure files
\usepackage{dcolumn}% Align table columns on decimal point
\usepackage{bm}% bold math

\def\cA{{\cal A}}

\newcommand{\req}[1]{Eq.~(\ref{#1})}
\newcommand{\avg}[1]{\langle #1\rangle}

\newcommand{\fig}[1]{Fig.~\ref{#1}}

\begin{document}
\title{Self-Organization of Balanced Nodes in Random
Networks with Transportation Bandwidths}
\subtitle{Self-Organization of Balanced Nodes}
\author{C. H. Yeung \and K. Y. Michael Wong % etc
% \thanks is optional - remove next line if not needed
%\thanks{\emph{Present address:} Insert the address here if needed}%
}                     % Do not remove
\offprints{}          % Insert a name or remove this line
\institute{Department of Physics, The Hong Kong University
of Science and Technology, 
Hong Kong, China}
\date{Received: date / Revised version: date}
% The correct dates will be entered by Springer
%
\abstract{
We apply statistical physics to study the task of resource allocation
in random networks with limited bandwidths along the transportation links.
The mean-field approach is applicable 
when the connectivity is sufficiently high. 
It allows us to derive
the resource shortage of a node as a well-defined function of its capacity.
For networks with uniformly high connectivity,
an efficient profile of the allocated resources is obtained, 
which exhibits features similar to the Maxwell construction.
These results have good agreements with simulations,
where nodes self-organize to balance their shortages,
forming extensive clusters of nodes interconnected by unsaturated links.
The deviations from the mean-field analyses
show that nodes are likely to be rich in the locality of gifted neighbors.
In scale-free networks,
hubs make sacrifice for enhanced balancing of nodes with low connectivity.
 % end of PACS codes
} %end of abstract

\titlerunning{Self-organization of Balanced Nodes}  % abbreviated title (for running head)
\authorrunning{C. H. Yeung and K. Y. M. Wong}   % abbreviated author list (for running head)

\maketitle
\section{Introduction}
\label{sec_introduction}

Analytical techniques developed in statistical physics 
have been widely employed in the analysis of complex systems 
in a wide variety of fields,
such as neural networks \cite{hertz,nishimori},
econophysical models \cite{challet}, 
and error-correcting codes \cite{nishimori,kabashima}.
Recently,
a statistical physics perspective was successfully applied to the problem
of resource allocation on sparse random networks \cite{wong2006,wong2007,yeung2009b}.
Resource allocation is a well known network problem in the areas of computer science
and operations management \cite{peterson,ho}.
It is relevant to applications such as load balancing in computer networks,
reducing Internet traffic congestion, and streamlining network flow
of commodities \cite{shenker,rardin}.

In networks with finite bandwidths,
the problem of resource allocation was studied in \cite{yeung2009}.
We derived an algorithm
which enable us to find the optimal solutions without the need
of a global optimizer.
The mean-field approach was applicable 
when the connectivity is sufficiently high.
It allows us to derive
the resource shortage of a node as a well-defined function of its capacity,
which corresponds to the optimized and initial resource relation.
For networks with uniformly high connectivity
we derived the profile of the allocated resources 
which exhibits features similar to the Maxwell construction.
We generalized the analysis to networks with arbitrary connectivity
and compared the modified Maxwell construction with numerical solutions.

In this paper, 
we focus on the self-organization of nodes 
in achieving a balanced environment.
The analytical results are compared with simulations,
where nodes self-organize to balance their shortages.
After defining the model in Section \ref{sec_model}, 
we introduce the chemical potentials in Section \ref{sec_analysis}. 
In Section \ref{sec_Maxwell} we review 
the theory of the Maxwell construction, 
which forms the basis for predicting 
the existence of clusters of balanced nodes. 
The emergence of balanced nodes correspond to the 
success in the uniform allocation of resources.
We compare in Section \ref{sec_cluster} the statistics of saturated and unsaturated links,
and show the existence of extensive balanced clusters.
The deviations of the simulation results from the mean-field analyses
show the dependence of final state on the locality
of gifted and ungifted clusters in Section \ref{sec_gifted}.
We compare in Section \ref{sec_scalefree} 
the fraction of balanced nodes 
in scale-free and regular networks
and examine the role of hubs in resource allocation.

\section{The Model}
\label{sec_model}

We consider a network with $N$ nodes, 
labelled $i\!=\!1,\dots,N$.
Each node $i$ is randomly connected to $c$ other nodes.
The connectivity matrix is given by $\cA_{ij}=1, 0$ for connected and unconnected node pairs respectively.
Each node $i$ has a capacity $\Lambda_i$ randomly drawn from a distribution
$\rho(\Lambda_i)$.
Positive and negative values of $\Lambda_i$ correspond to supply and demand
of resources respectively.
The task of resource allocation involves transporting resources between
nodes such that the demands of the nodes can be
satisfied to the largest extent.
Hence we assign $y_{ij}\equiv-y_{ji}$ to be the {\it current} drawn from node $j$ to $i$,
aiming at reducing the {\it shortage} $\xi_i$ of node $i$ defined by
\begin{eqnarray}
\label{xi_define}
	\xi_i=\max\biggl(-\Lambda_i-\sum_{(ij)}\cA_{ij}y_{ij}, 0\biggr).
\end{eqnarray}
The magnitudes of the currents are bounded by the {\it bandwidth} $W$,
i.e., $|y_{ij}|\leq W$.

To minimize the shortage of resources after their allocation,
we include in the total cost both the shortage cost and the transportation cost.
Hence,
the general cost function of the system can be written as 
\begin{eqnarray}
\label{E_define}
	&&E=R\sum_{(ij)}\cA_{ij}\phi(y_{ij})
	+\sum_i\psi(\Lambda_i,\{y_{ij}|\cA_{ij}=1\}).
\end{eqnarray}
The summation $(ij)$ corresponds to summation over all node pairs,
and $\Lambda_i$ is a quenched variable defined on node $i$.

In the present model of resource allocation, 
the first and second terms correspond to the transportation and shortage 
costs respectively.
The parameter $R$ corresponds to the {\it resistance} on the currents,
and $\Lambda_i$ is the capacity of node $i$.
The transportation cost $\phi(y_{ij})$ can be a general even function
of $y_{ij}$.
In this paper, 
we consider $\phi$ and $\psi$ to be concave functions of their arguments,
that is,
$\phi'(y)$ and $\psi'(\xi)$ are non-decreasing functions.
Specifically, 
we have the quadratic transportation cost $\phi(y)=y^2/2$, 
and the quadratic shortage cost $\psi(\Lambda_i, \{y_{ij}|\cA_{ij}=1\})=\xi_i^2/2$.

%%%%%%%%%%%%%%%%%%%%%%%%%%%%%%%%%%%%%%%%%%%%%%%%%
\section{The Chemical Potentials and the Final Resources}
\label{sec_analysis}

The optimization problem 
can be written as the minimization of \req{E_define} in the space of 
$y_{ij}$ and $\xi_i$,
subject to the constraints
\begin{eqnarray}
\label{xiCon}
	\Lambda_i+\sum_{(ij)}\cA_{ij}y_{ij}+\xi_i\ge 0 ,
	\quad\quad
	\xi_i\ge 0 ,
\end{eqnarray}
and the constraints on the bandwidths of the links $|y_{ij}|\le W$.
Introducing Lagrange multipliers to the above inequality constraints
with the Kuhn-Tucker condition,
the function to be minimized becomes 
\begin{eqnarray}
\label{Lagr}
	L\!=\!\sum_i\biggl[\psi(\xi_i)+\mu_i\biggl(\Lambda_i+\sum_{(ij)}\cA_{ij}y_{ij}+\xi_i\biggr)
	+\alpha_i\xi_i\biggr]
	\nonumber\\
	+\sum_{(ij)}\cA_{ij}\biggl[R\phi(y_{ij})+\gamma^+_{ij}(W-y_{ij})+\gamma^-_{ij}(W+y_{ij})\biggr],
\end{eqnarray}
where $\mu_i\leq 0$, $\alpha_i\leq 0$, $\gamma_{ij}^+\leq 0$ and 
$\gamma_{ij}^-\leq 0$.
Optimizing $L$ with respect to $y_{ij}$, 
one obtains
\begin{eqnarray}
\label{solution}
	y_{ij} =Y(\mu_j-\mu_i)
\end{eqnarray}
with
\begin{eqnarray}
\label{solutionY}
	Y(x) = \max\biggl\{-W, \min\biggl[W, [\phi']^{-1}\biggl(\frac{x}{R}\biggr)\biggr]\biggr\}.
\end{eqnarray}
The Lagrange multiplier $\mu_i$ is referred to as the {\it chemical potential}
of node $i$,
and $\phi'$ is the derivative of $\phi$ with respect to its argument.
The function $Y(\mu_j-\mu_i)$ relates the potential difference between
nodes $i$ and $j$ to the current driven from node $j$ to $i$.
For the quadratic cost,
it consists of a linear segment between $\mu_j-\mu_i = \pm WR$ reminiscent of Ohm's law
in electric circuits.
Beyond this range,
$y$ is bounded above and below by $\pm W$ respectively.
Thus,
obtaining the optimized configuration of currents $y_{ij}$ among the nodes 
is equivalent to finding the corresponding set of chemical potentials $\mu_i$,
from which the optimized $y_{ij}$'s are then derived from $Y(\mu_j-\mu_i)$.
This implies that we can consider the original optimization problem
in the space of chemical potentials.

The optimal currents are given by \req{solution} in terms of the chemical potentials
$\mu_i$ which, 
from Eqs. (\ref{xi_define}) and (\ref{Lagr}),
are related to their neighbors via 
\begin{eqnarray}
\label{CPmu}
	\mu_{i}=
	\begin{cases}
	0 \qquad\qquad\qquad\mbox{ for $h_{i}^{-1}(0)>0$,}
	\\
	h_{i}^{-1}(0) \qquad\qquad\mbox{ for $-\psi'(0)\leq h_{i}^{-1}(0)\leq 0$,}
	\nonumber\\ 
	g_{i}^{-1}(0) \qquad\qquad\mbox{ for $ h_{i}^{-1}(0)< -\psi'(0)$,}
	\end{cases}
	\\
\end{eqnarray}
where $h_i(x)$ and $g_i(x)$ are given by 
\begin{eqnarray}
\label{CPhg}
	h_i(x) &=& -\Lambda_i-\sum_j\cA_{ij}Y(\mu_j-x),
	\nonumber\\
	g_i(x) &=& \psi'\circ h_i(x) + x,
\end{eqnarray}
with function $Y$ again given \req{solutionY}.
$h_i(x)$ is the shortage of resource at node $i$ when $\mu_i$ takes the value $x$.
$\psi'\circ h_i(x)$ is then the corresponding dissatisfaction cost per unit resource
of node $j$. 
For the quadratic shortage cost considered in this paper, 
the {\it frictionless} condition $\psi'(0)=0$ is satisfied. 
Equation~(\ref{CPmu}) is then simplified to
\begin{equation}
\label{CPmu2}
	\mu_{i}=\min\left(0,-h_i(\mu_i)\right).
\end{equation}
Hence we can interpret $\mu_i$ as the final shortage of resources 
after optimization. 
When $\mu_i<0$, $-\mu_i$ becomes the final resources allocated to node $i$. 
Equation~(\ref{CPmu2}) provides a simple local iteration algorithm 
for the optimization problem 
in which the optimal currents can be evaluated from the potential
differences of neighboring nodes.

An alternative algorithm can be obtained by adopting
message-passing approaches, 
which have been successful in problems such as error-correcting codes
\cite{opper2001} and probabilistic inference \cite{mackay2003}.
We refer the interested readers to \cite{yeung2009} for 
a comprehensive derivation of the messages.

%%%%%%%%%%%%%%%%%%%%%%%%%%%%%%%%%%%%%%%%%%%%%%%%%
\section{The Resource Distribution Profile}
\label{sec_Maxwell}

\subsection{The High Connectivity Limit}
\label{sec_highC}

We consider the case that the bandwidth of individual links scales as 
$\tilde W/c$ when the connectivity increases,
where $\tilde W$ is a constant. 
Thus the total bandwidth $\tilde W$ available to an individual node remains 
a constant.

%%%%%%%%%%%%%%%%%%%%%%%%%%%%%%
%\subsubsection{The Well-defined Chemical Potential Function}

We start by writing the chemical potentials using \req{CPmu2},
\begin{eqnarray}
\label{vbmu}
	\mu_i=\min\biggl[\Lambda_i
	+\sum_{j=1}^N\cA_{ij}Y(\mu_j-\mu_i),0\biggr].
\end{eqnarray}
In the high connectivity limit,
the interaction of a node with all its connected neighbors become self-averaging, 
making it a function singly dependent on its own chemical potential,
namely,
\begin{eqnarray}
\label{vbMmu}
	\sum_{j=1}^N\cA_{ij}Y(\mu_j-\mu_i)\approx c M(\mu_i).
\end{eqnarray}
Physically,
the function $M(\mu)$ corresponds to the average interaction of a node
with its neighbors when its chemical potential is $\mu$, 
facilitating a mean-field approach. 
Thus,
we can write Eq.~(\ref{vbmu}) as
\begin{eqnarray}
\label{vbmu1a}
	\mu=\min[\Lambda+c M(\mu),0],
\end{eqnarray}
where $\mu$ is now a function of $\Lambda$, and
\begin{eqnarray}
\label{vbMmuint1}
	M(\mu_i)
	=\int_{-\infty}^\infty d\Lambda\rho(\Lambda)
	Y(\mu(\Lambda)-\mu_i)
\end{eqnarray}
where we have written the chemical potential of the neighbors as $\mu(\Lambda)$,
assuming that they are well-defined functions of their capacities $\Lambda$.

To explicitly derive $M(\mu)$, 
we take advantage of the fact that the rescaled bandwidth, 
$\tilde W/c$ vanishes in the high connectivity limit,
so that the current function $Y(\mu_j-\mu_i)$ is effectively a sign function, 
corresponding to saturated links. 
(This approximation is not fully valid 
and will be further refined in subsequent discussions.)
Thus, 
we approximate
\begin{eqnarray}
\label{vbMmuint2}
	M(\mu_i)
	=\frac{\tilde W}{c}\int_{-\infty}^\infty d\Lambda\rho(\Lambda)
	{\rm sgn}[\mu(\Lambda)-\mu_i].
\end{eqnarray}
Assuming that $\mu(\Lambda)$ is a monotonic function of $\Lambda$,
and for Gaussian distribution of capacities, 
$\mu(\Lambda)$ is explicitly given by 
\begin{eqnarray}
\label{vbmu2}
	\mu=\min\biggl[\Lambda-\tilde W{\rm erf}\biggl(
	\frac{\Lambda-\langle\Lambda\rangle}{\sqrt{2}}\biggr),0\biggr].
\end{eqnarray}
This equation relates the chemical potential of a node, 
i.e. the shortage after resource allocation, 
to its initial resource before.
It tells us that resource allocation through a large number of links results 
in a well-defined function relating the two quantities.

%\subsubsection{The Maxwell Construction}

\begin{figure}
\centerline{\epsfig{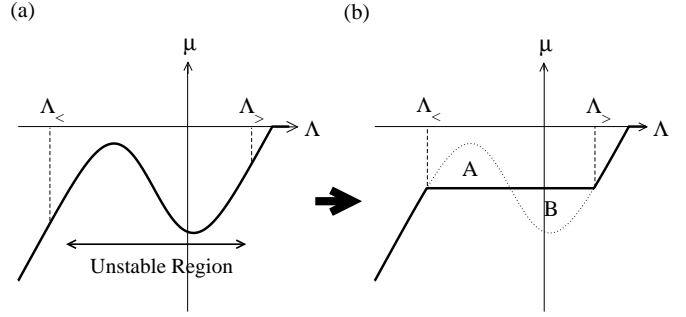}}
\caption{The Maxwell construction on $\mu(\Lambda)$.}
\label{gr_maxwell}
\end{figure}

\req{vbmu2} gives a well-defined function $\mu(\Lambda)$ as long as
$\tilde W\leq \sqrt{\pi/2}$.
However,
when $\tilde W> \sqrt{\pi/2}$,
turning points exists in $\mu(\Lambda)$ as shown in Fig.~\ref{gr_maxwell}(a).
This creates a thermodynamically unstable scenario,
since in the region of $\mu(\Lambda)$ with negative slope,
nodes with lower capacities have higher chemical potentials than their neighbors with 
higher capacities.
Mathematically,
the non-monotonicity of $\mu(\Lambda)$ means that ${\rm sgn}[\mu(\Lambda)-\mu_i]$ 
and ${\rm sgn}(\Lambda-\Lambda_i)$ are no longer necessarily equal,
and \req{vbmu2} is no longer valid.

%%%%%%%%Figure1%%%%%%%%%%%%%%%%%%%%
%%%%%%%%%%%%%%%%%%%%%%%%%%%%%%%%%

Nevertheless,
\req{vbmu} permits another solution of constant $\mu$ in a range of $\Lambda$.
Hence,
we propose that the unstable region of $\mu(\Lambda)$ should be 
replaced by a range of constant $\mu$ as shown in Fig.~\ref{gr_maxwell}(b)
analogous to the Maxwell construction in thermodynamics. 
Nodes within this range of constant $\mu$ 
have the same amount of final resources, 
and is the consequence of the ability of the optimization process 
to balance the resources. 
They are referred to as the {\it balanced} nodes.

In the high connectivity limit,
resources are so efficiently allocated that the resources of the rich nodes
are maximally allocated to the poor nodes.
By considering the conservation of resources, 
and letting $(\Lambda_<, \mu_o)$ and $(\Lambda_>, \mu_o)$ be the end points of 
the Maxwell construction as shown in Fig.~\ref{gr_maxwell}(b),
we have proved in \cite{yeung2009} that
\begin{eqnarray}
	\mu_o\int_{\Lambda_<}^{\Lambda_>} d\Lambda\rho(\Lambda)
	= \int_{\Lambda_<}^{\Lambda_>} d\Lambda\rho(\Lambda)\mu(\Lambda),
\end{eqnarray}
which implies that the value of $\mu_o$ should be chosen such that
the areas A and B in Fig.~\ref{gr_maxwell}(b),
weighted by the distribution $\rho(\Lambda)$,
should be equal.

For capacity distributions $\rho(\Lambda)$ symmetric with respect to $\avg{\Lambda}$,
we have $\mu_o = \avg{\Lambda} = (\Lambda_<+\Lambda_>)/2$.
As a result, 
the function $\mu(\Lambda)$ is given by 
\begin{eqnarray}
\label{vbmuhori}
	\mu(\Lambda) = 
	\begin{cases}
  	\langle\Lambda\rangle   \qquad\qquad\qquad\qquad\mbox{ for $\Lambda_<<\Lambda<\Lambda_>$,}\\
  	\\
  	\min\left[\Lambda-
  	\tilde W{\rm erf}\left(\frac{\Lambda-\langle\Lambda\rangle}{\sqrt{2}}\right),0\right]
  	\mbox{ otherwise,}\\
	\end{cases} 
\end{eqnarray}
where as $\Lambda_<$ and $\Lambda_>$ are respectively given by the lesser and 
greater roots of the equation $x=\langle\Lambda\rangle
+\tilde W{\rm erf}[(x-\langle\Lambda\rangle)/\sqrt{2}]$.

%%%%%%%% Figure2 %%%%%%%%%%%%%%%%%%%%
\begin{figure}
%\vspace{-20pt}
\centerline{\epsfig{figure=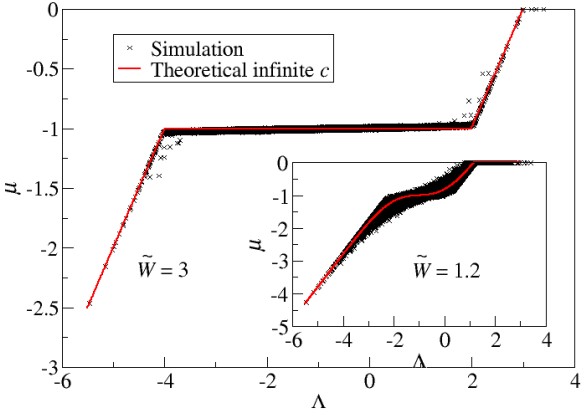, width=\linewidth}}
\caption{The simulation results of $\mu(\Lambda)$ 
for $N=10000$, $c=15$, $R=0.1$, $\langle\Lambda\rangle = -1$ and $\tilde W=3$
with 70000 data points, 
compared with theoretical prediction.
Inset: The corresponding results for $\tilde W=1.2$.}
\label{gr_vbhori}
\end{figure}
%%%%%%%%%%%%%%%%%%%%%%%%%%%%%%%%%

We compare the analytical result of $\mu(\Lambda)$ in Eq.~(\ref{vbmuhori})
with simulations 
in Fig.~\ref{gr_vbhori}.
For $\tilde W > \sqrt{\pi/2}$, 
data points $(\Lambda, \mu)$ of individual nodes from network simulations 
follow the analytical result
of $\mu(\Lambda)$, 
giving an almost perfect overlap of data.
The presence of the balanced nodes with effectively constant chemical potentials 
is obvious and essential to explain the behavior of the majority of data
points from simulations.
On the other hand,
for $\tilde W<\sqrt{\pi/2}$, 
the analytical $\mu(\Lambda)$ shows no turning point as shown in 
the inset of Fig.~\ref{gr_vbhori}.
Despite the scattering of data points,
they generally follow the trend 
of the theoretical $\mu(\Lambda)$.

%%%%%%%%%%%%%%%%%%%%%%%%%%%%%%%%%%%%%%%%%%
\subsection{The Cases with General Connectivity}

Our analysis can be generalized to the case of large but finite connectivity, 
where the approximation in \req{vbMmuint2} is not fully valid.
This modifies the chemical potentials of the balanced nodes,
for which \req{vbMmuint2} has to be replaced by 
\begin{eqnarray}
\label{vbslantMmu}
	M(\mu)&&=
	\frac{\tilde W}{c}\biggl[
	\int_{\Lambda_>}^{\infty}d\Lambda\rho(\Lambda)
	-\int_{-\infty}^{\Lambda_<}d\Lambda\rho(\Lambda)\biggr]
	\nonumber\\
	&&+\int_{\Lambda_<}^{\Lambda_>}d\Lambda\rho(\Lambda)
	\biggl(\frac{\mu(\Lambda)-\mu}{R}\biggr).
\end{eqnarray}
We introduce an ansatz of a linear relationship between
$\mu$ and $\Lambda$ for the balanced nodes,
namely,
\begin{eqnarray}
\label{vbslantanastz}
	\mu=m\Lambda+b.
\end{eqnarray}
After direct substitution of \req{vbslantanastz} into $M(\mu)$ given by 
\req{vbslantMmu},
we get the self-consistent equations for $m$ and $b$,
\begin{eqnarray}
\label{vbslantmb}
	m=\frac{R}{R+c~{\rm erf}\left(
	\frac{\Lambda_>-\langle\Lambda\rangle}{\sqrt{2}}\right)},
	\nonumber\\
	b=\frac{c~{\rm erf}\left(
	\frac{\Lambda_>-\langle\Lambda\rangle}{\sqrt{2}}\right)}
	{R+c~{\rm erf}\left(
	\frac{\Lambda_>-\langle\Lambda\rangle}{\sqrt{2}}\right)}\langle\Lambda\rangle.
\end{eqnarray}
Thus,
the Maxwell construction has a non-zero slope when the connectivity is finite.

We remark that the approximation in \req{vbslantMmu} assumes that the 
potential differences of the balanced nodes lie in the range of $2R\tilde W/c$,
so that their connecting links remain unsaturated.
Note that the end points of the Maxwell construction have chemical potentials
$\avg{\Lambda}\pm R\tilde W/c$ respectively,
rendering the approximation in \req{vbslantMmu} {\it exact} at one special point,
namely,
the central point of the Maxwell construction. 
Hence,
this approximation works well in the central region of the Maxwell construction,
while deviations are expected near the end points.

\begin{figure}
\centerline{\epsfig{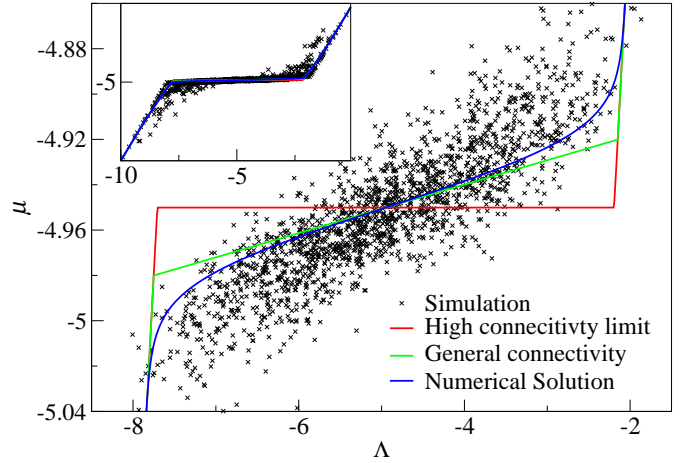}}
\caption{
Simulation results of $(\Lambda, \mu)$
for $N=2000$, $\tilde W=3$, $c=10$ and $\langle\Lambda\rangle=-5$ 
shown with an expanded vertical scale,
as compared with the theoretical predictions of $\mu(\Lambda)$
from \req{vbmuhori}, \req{vbslantMmu} and the numerical solution of 
\req{eq_ns}. 
Inset: same data set and theoretical predictions 
in the normal vertical scale.
}
\label{gr_slant}
\end{figure}

We compare \req{vbslantMmu} with the $\mu(\Lambda)$
given by the numerical solution of the integral equation
\begin{eqnarray}
\label{eq_ns}
	\mu(\Lambda_i) = \Lambda_i 
	+\int_{-\infty}^{\infty}d\Lambda \rho(\Lambda)Y[\mu(\Lambda)-\mu(\Lambda_i)].
\end{eqnarray}
Since iterating this equation may lead to an oscillating solution of $\mu(\Lambda)$, 
we solve it by gradient descent. 
The results are shown in \fig{gr_slant} in an enlarged scale of $\mu$. 
As expected, 
\req{vbslantMmu} works well around $\mu=\avg{\Lambda}$
and show small deviations at the end points of the Maxwell construction.
In comparison with simulations,
data porints are scattered from the theoretical predictions,
but generally follow the slanted path of $\mu(\Lambda)$
rather than the horizontal path as predicted by \req{vbmuhori}.
We will explain the scattering of data points in the next section.
As shown in the inset of \fig{gr_slant},
the differences among the different approaches
are not obvious unless the scale of $\mu$ is expanded. 
We thus conclude that both the horizontal and slanted Maxwell constructions 
are good approximations of $\mu(\Lambda)$
and capture the general trend of the simulation data.

%%%%%%%% Figure3 %%%%%%%%%%%%%%%%%%%%
%%%%%%%%%%%%%%%%%%%%%%%%%%%%%%%%%

Remarkably, 
as evident from Eq.~(\ref{vbslantmb}),
even with constant available bandwidth $\tilde W$, 
increasing connectivity causes $m$ to decrease,
and hence sharpens the chemical potential distribution.
The narrower distributions correspond to higher efficiency in resource allocation.
It leads us to realize the potential benefits of 
increasing connectivity in network
optimization even for a given constant total bandwidth connecting a node.

%%%%%%%%%%%%%%%%%%%%%%%%%%%%%%%%%%%%%%%
\section{The Self-organization of the Balanced Nodes}
\label{sec_balanced}

\begin{figure}
\centerline{\epsfig{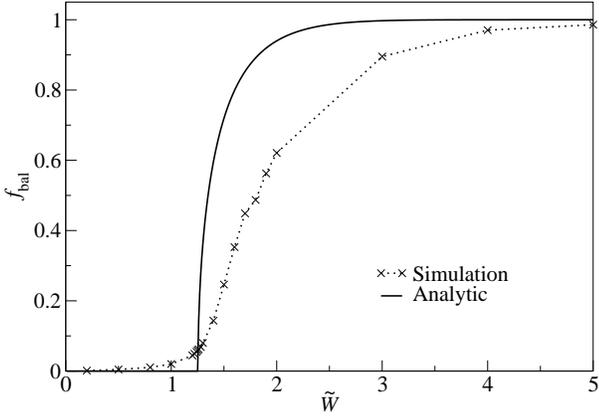}}
\caption{
Analytical predictions of $f_{\rm bal}$ as compared with
simulation results 
for networks of $N=10000$, $\tilde W=3$, $R=0.1$ and $\langle\Lambda\rangle=-5$.
}
\label{gr_fbal}
\end{figure}

The fraction $f_{\rm bal}$ of balanced nodes is given by the equation
\begin{eqnarray}
\label{eqfbal}
	f_{\rm bal}={\rm erf}\biggl(\frac{\tilde W f_{\rm bal}}{\sqrt{2}}\biggl).
\end{eqnarray}
Note that $f_{\rm bal}$ has the same dependence on $\tilde W$ for all negative $\avg{\Lambda}$.
Figure \ref{gr_fbal} shows that when
the total bandwidth $\tilde W$ increases beyond $\sqrt{\pi/2}$,
the analytical fraction of balanced nodes increases,
reflecting the more efficient resource allocation brought by the 
convenience of increased bandwidths.
When $\tilde W$ becomes very large,
a uniform chemical potential of $\avg{\Lambda}$ networkwide is recovered. 
converging to the case of non-vanishing bandwidths \cite{yeung2009}.

%%%%%%%% Figure4 %%%%%%%%%%%%%%%%%%%%
%%%%%%%%%%%%%%%%%%%%%%%%%%%%%%%%%

We measure $f_{\rm bal}$ in simulations as follows.
As only finite connectivity can be implemented,
we define node $i$ to be balanced when its chemical potential 
falls into the slanted range of the Maxwell construction,
i.e. $\avg{\Lambda}-R\tilde W/c\le\mu_i\le\avg{\Lambda}+R\tilde W/c$.
The simulation results are compared with the analytical results
in \fig{gr_fbal}.
Deviations are found at intermediate values of $\tilde W$,
which may be expained by the scattering of simulated data points.
Nevertheless,
increass in $f_{\rm bal}$ are observed at $\tilde W\approx \sqrt{\pi/2}$,
corresponding to the emergence of balanced nodes in simulations.

\subsection{The Extensive Clusters of Balanced Nodes}
\label{sec_cluster}

In random networks,
balanced nodes are found in clusters
interconnected by an extensive fraction of unsaturated links.
The clusters connect most balanced nodes 
and span the whole network when 
a large fraction of balanced node is found.
The unsaturated links in the clusters
provide the freedom to fine tune their currents so that
the shortages among the nodes are uniform.
These features are the natural consequences of optimization,
in which nodes self-organize to balance their shortages.

%%%%%%%% Figure4 %%%%%%%%%%%%%%%%%%%%
\begin{figure*}
\leftline{\epsfig{figure=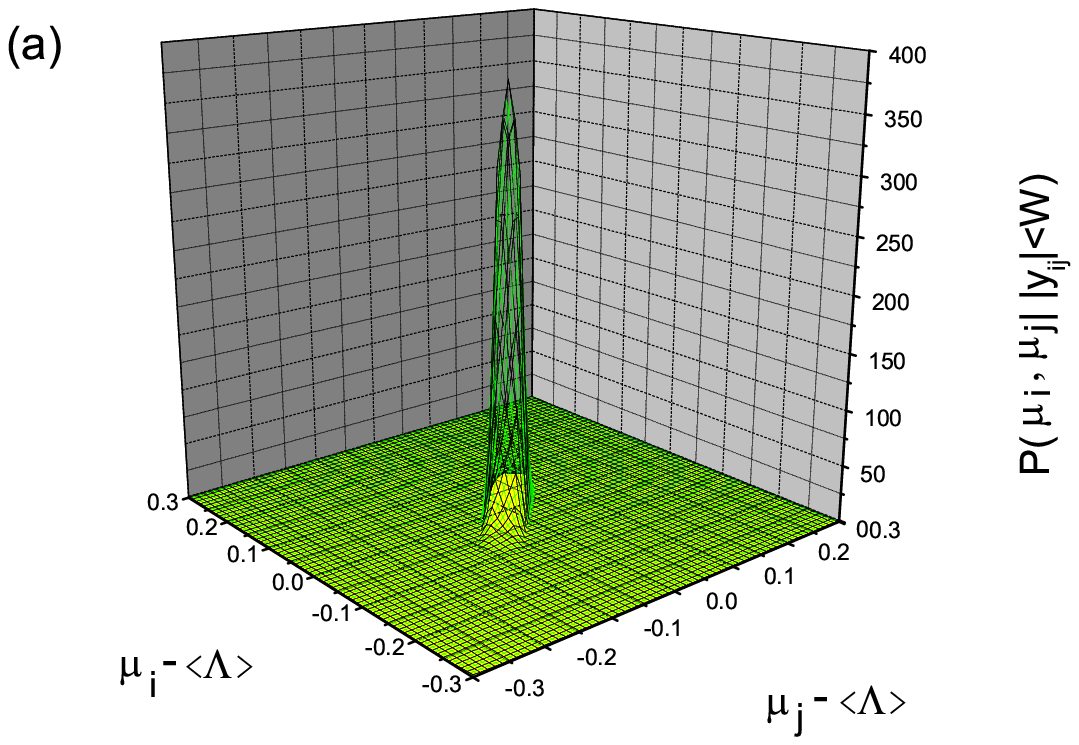, width=0.48\linewidth}
\leftline{\epsfig{figure=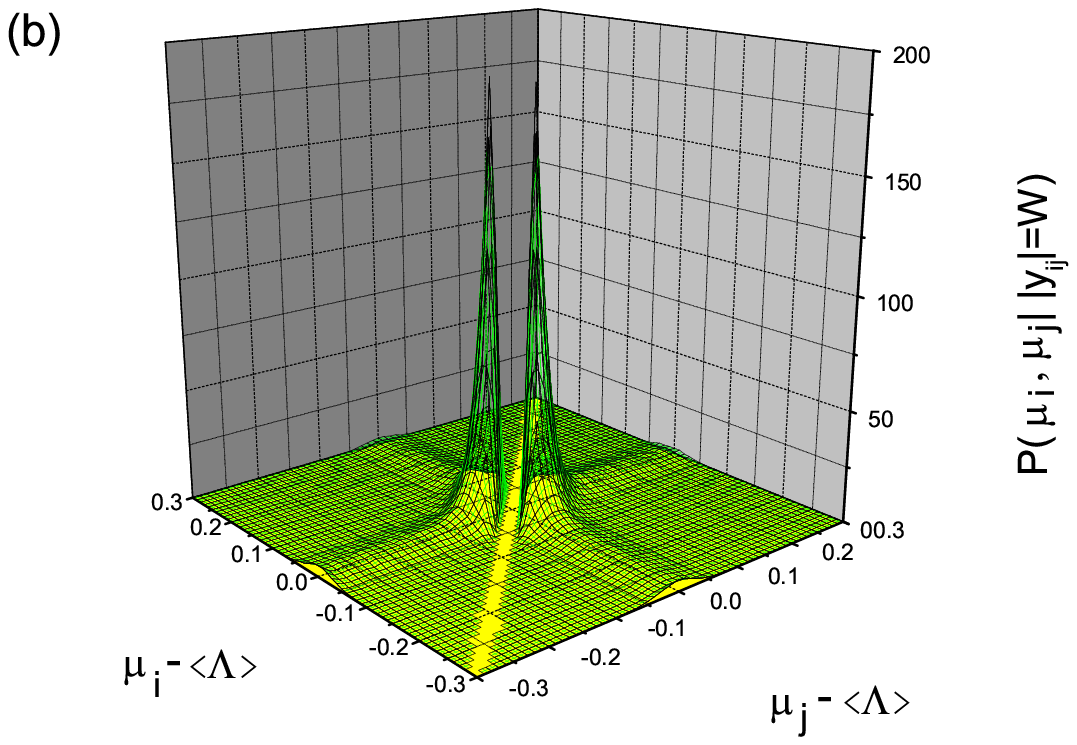, width=0.48\linewidth}}
}
\caption{
The distributions (a) $P(\mu_i, \mu_j| |y_{ij}|<W)$
and (b) $P(\mu_i, \mu_j| |y_{ij}|=W)$,
corresponding to the joint probability distributions of 
the final resources of terminal nodes $i$ and $j$ for link $(ij)$,
given the link is unsaturated and saturated respectively. 
Parameters: $N=10000$, $\tilde W=3$, $R=0.1$ and $\langle\Lambda\rangle=-5$.
}
\label{gr_condP}
\end{figure*}
%%%%%%%%%%%%%%%%%%%%%%%%%%%%%%%%%

To examine the clustering of balanced nodes,
we show in \fig{gr_condP} (a) and (b)
respectively the distributions $P(\mu_i, \mu_j| |y_{ij}|<W)$
and $P(\mu_i, \mu_j| |y_{ij}|=W)$,
which correspond to the joint probability distributions of 
the final resources of terminal nodes $i$ and $j$ for link $(ij)$,
given the link is unsaturated and saturated.
As shown in \fig{gr_condP} (a),
unsaturated links connect nodes with $\mu_i\approx\mu_j$.
A prominent peak is found around $\mu_i\approx\mu_j\approx\avg{\Lambda}$,
corresponding to unsaturated linkages between the balanced nodes.
On the other hand,
$P(\mu_i, \mu_j| |y_{ij}|=W)$ shows non-zero probabilities 
in regions other than $\mu_i\approx\mu_j$.
Peaks are observed,
which are similar to a Gaussian distribution with 
the central slice removed. 
Non-zero probabilities are observed along the axes 
$\mu_i\approx\avg{\Lambda}$ and $\mu_j\approx\avg{\Lambda}$,
corresponding to saturated linkages between balanced and unbalanced nodes.
These results support the existence of balanced clusters 
interconnected by an extensive fraction of unsaturated links.

\subsection{The Neighborhood of Rich and Poor Nodes}
\label{sec_gifted}

\begin{figure}
\centerline{\epsfig{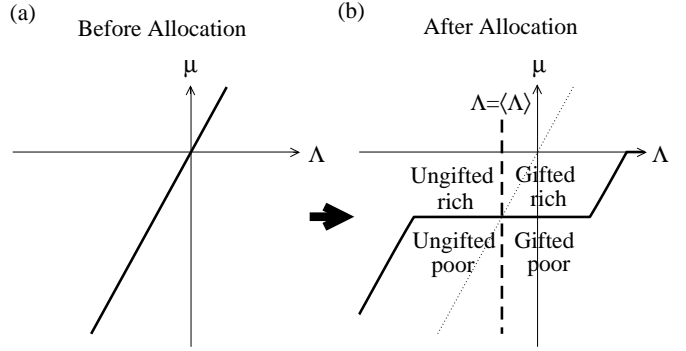}}
\caption{
The schematic relation of the resources of a node 
(a) before and (b) after optimization.
The division of nodes into gifted rich,
gifted poor, ungifted rich and ungifted poor is shown in (b).}
\label{gr_category}
\end{figure}

To understand the scattering of data points of $(\mu, \Lambda)$
from the mean-field predictions, 
we identify the role of the nodes in resource allocation 
according to their capacities.
Nodes with capacities 
greater and less than $\avg{\Lambda}$ are respectively referred to as the
{\it gifted} and {\it ungifted} nodes. 
Figure \ref{gr_category} shows the schematic relation of the resources of a node 
before and after optimization. 
Before optimization, 
the resource of a node is equal to its capacity.
After optimization, 
the resource of a node is equal to $\mu$. 
Gifted nodes have their resources reduced 
after donating them, 
and ungifted nodes have their resources increased 
after receiving them. 
$\mu(\Lambda)$ is then described by the Maxwell construction 
as derived in the mean-field analysis.

%%%%%%%%Figure6%%%%%%%%%%%%%%%%%%%%
%%%%%%%%%%%%%%%%%%%%%%%%%%%%%%%%%

As finite connectivity is implemented in simulations,
the neighborhood of a node deviates from the mean-field descriptions
which results in scattering of data points $(\mu, \Lambda)$.
To examine the effect of the locality of nodes in relation with 
their final shortage, 
we define 
{\it rich} and {\it poor} nodes to be nodes with final resources
higher and lower than the mean-field predictions.
As shown in \fig{gr_category}(b),
the nodes are thus categorized into 
{\it gifted rich}, {\it gifted poor}, {\it ungifted rich}
and {\it ungifted poor},
in accordance to the scattering of the
capacity-shortage relations
i.e. $(\mu, \Lambda)$ of the node.
As an example,
gifted rich nodes are nodes with initial capacity  
higher than $\avg{\Lambda}$ 
and final resources higher than the mean-field predictions,
as shown in \fig{gr_category}(b).

%%%%%%%%Figure6%%%%%%%%%%%%%%%%%%%%
\begin{figure}
\centerline{\epsfig{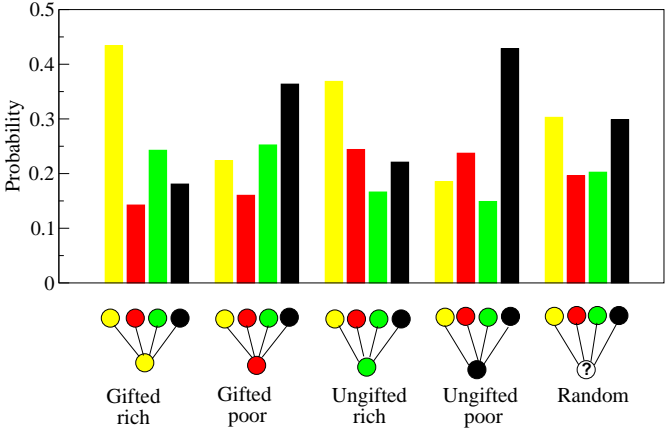}}
\caption{
The fraction of gifted rich, gifted poor, 
ungifted rich and ungifted poor
nodes among the nearest neighbors
of a node in a particular category.
Parameters:  $N=10000$, $\tilde W=3$, $R=0.1$, $\langle\Lambda\rangle=-5$
and 5 samples.
}
\label{gr_barChart}
\end{figure}
%%%%%%%%%%%%%%%%%%%%%%%%%%%%%%%%%

We examine the neighborhood of a node in \fig{gr_barChart} by measuring 
the fractions of gifted rich, gifted poor, ungifted rich and ungifted poor
nodes among its nearest neighbors.
As compared with the random case,
a high ratio of gifted rich node is found
surrounding a gifted rich node.
In other words,
gifted nodes are more likely to be rich 
in the neighborhood of gifted nodes,
forming a cluster of rich nodes after allocation.
Physically,
the effective average capacity 
is higher than $\avg{\Lambda}$ in the locality of gifted clusters,
leading to higher resources than the mean-field predictions.
The converse is true for ungifted poor nodes,
which results in lower resources in the locality of ungifted clusters.
On the other hand, 
gifted nodes are more likely to be poor 
if they are in the neighborhood of ungifted clusters,
as shown by the statistics of the neighbors of gifted poor nodes,
and vice versa.
We thus conclude that the final state of a node is highly 
dependent on its locality,
which results in the scattering of the simulated data points of $(\mu, \Lambda)$
around the prediction of the mean-field analyses.

%%%%%%%%%%%%%%%%%%%%%%%%%%%%%%%%%%%%%%%%%%%%%%%
\section{Balanced Nodes in Scale-Free Networks}
\label{sec_scalefree}

%%%%%%%% Figure4 %%%%%%%%%%%%%%%%%%%%
\begin{figure}
\centerline{\epsfig{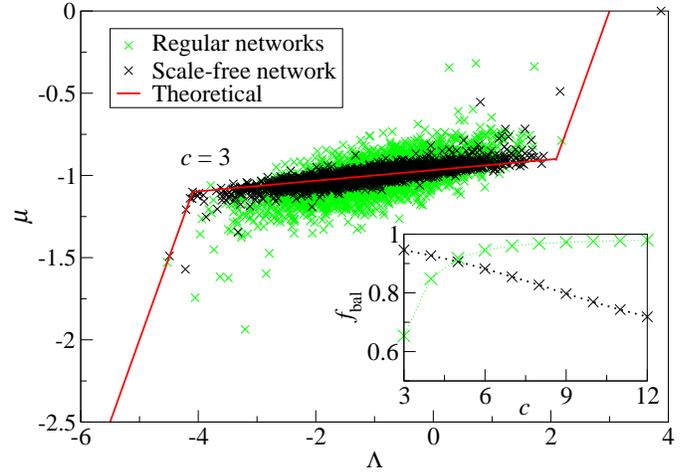}}
\caption{Simulation results of $(\Lambda, \mu)$
for networks of $N=2\times10^5$, $\tilde W=3$, $R=0.1$ and $\langle\Lambda\rangle=-1$
with uniform connectivity of $c=3$ and
scale-free network of $P(c)\sim c^{-3}$ with $c\ge 3$, 
each with 2500 data points, 
as compared with the theoretical predictions of Eq. (\ref{vbslantmb}). 
Inset: the $f_{\rm bal}$ as a function of $c$ for nodes in 
scale-free and regular networks with N=10000 and 50 samples.
}
\label{figScaleFree}
\end{figure}
%%%%%%%%%%%%%%%%%%%%%%%%%%%%%%%%%

We have examined the features of balanced nodes in regular networks .
However,
recent studies of complex networks show that many realistic communication 
networks have highly heterogeneous structure,
and the connectivity distribution obeys a power law \cite{barabasi}.
These networks,
commonly known as scale-free networks,
are characterized by the presence of hubs,
which are nodes with very high connectivities,
and are found to modify the network behavior significantly.
Hence,
it is interesting to study the allocation of resources 
and the features of balanced nodes in scale-free networks.
We define the bandwidth of the link $(ij)$ to be $W_{ij}=\tilde W/\max(c_i, c_j)$,
where $c_i$ and $c_j$ are the connectivity of the terminal nodes.
In this case,
nodes in scale-free network may have a smaller effective $\tilde W$,
as compared with their counterpart in regular networks with identical connectivity.

The simulation results are presented in Fig. \ref{figScaleFree},
where we plot the data points of $(\Lambda, \mu)$ from nodes of $c=3$ in 
scale-free networks.
Despite their low connectivity,
their capacity-shortage relations exhibit the flat distribution 
characteristic of the Maxwell construction,
coinciding with the analytical results of the high
connectivity limit.
This shows that the presence of hubs in scale-free networks increases the 
global efficiency of resource allocation,
leading to balanced shortages on nodes with low connectivity.
To confirm this advantage of the scale-free topology,
we also plot in the figure the data points obtained from networks
of uniform connectivity $c=3$.
Evidently,
the data points are much more scattered away from the Maxwell construction.

However,
the enhanced balancing in scale-free networks 
are found only for nodes with low connectivity.
We compare in the inset of Fig. \ref{figScaleFree},
the $f_{\rm bal}$ in scale-free networks
and regular networks,
for nodes with higher connectivities.
From the firgure for $c=3,4$,
a much higher $f_{\rm bal}$ is found 
in scale-free networks than their counterparts
in regular networks.
The opposite is true for $c\ge 5$,
and the differences increases with $c$.
It implies that the nodes with higher connectivity in scale-free neworks sacrifice
themselves for balancing the majority of nodes with low connectivity.
In contrast, the fraction of balanced nodes increases with the connectivity
in regular networks.
This picture thus clarifies the role of hubs in resource allocation on 
scale-free networks.

%%%%%%%%%%%%%%%%%%%%%%%%%%%%%%%%%%%%%%%%%%%%%%%%%%%%%
\section{Conclusion}
\label{sec:Conclusion}

We have applied statistical mechanics 
to study an optimization task of resource allocation on a network,
in which nodes with different capacities are connected 
by links of finite bandwidths.
By adopting suitable cost functions, 
such as quadratic transportation and shortage costs,
the model can be applied to the study of realistic networks.
The mean-field approach valid in the high connectivity limit 
enables us to derive the capacity-shortage relations, 
and study the deviations from this limit for finite connecitivty.

In particular,
the study reveals interesting effects due to finite bandwidths.
A remarkable phenomenon is found in networks with fixed total bandwidths per node,
where bandwidths per link vanish in the high connectivity limit.
For sufficiently large total bandwidths,
clusters of balanced nodes self-organize to have a uniform shortage
reminiscent of the Maxwell construction in thermodynamics.
The locality of gifted and ungifted clusters respectively lead to 
the formation of rich and poor clusters.
In scale-free networks, 
hubs are more likely to be unbalanced
and make sacrifice for nodes with
low connectivity to get balanced. 
We believe that the present analyses of balanced nodes lead us to better understanding
of self-organization in resource allocation,
as well as other systems.

%%%%%%%%%%%%%%%%%%%%%%%%%%%%%%%%%%%%%%%%%%%%%%%%%%%%%%%%%%%%%%%%%%
\section*{Acknowledgements}
This work is supported
by the Research Grant Council of Hong Kong
(grant numbers HKUST 603607 and HKUST 604008).

%
% BibTeX users please use
% \bibliographystyle{}
% \bibliography{}
%
% Non-BibTeX users please use

\end{document}